\documentstyle[aps,epsf,floats,twocolumn,dcolumn]{revtex}
\newcommand{\be}{\begin{eqnarray}}
\newcommand{\ee}{\end{eqnarray}}
\newcommand{\beq}{\begin{equation}}
\newcommand{\eeq}{\end{equation}}
\begin{document}
\wideabs{
\begin{flushright}
JLAB-THY-01-16 \\
June 5, 2001 \\
hep-ph/0106059
\end{flushright}
\title{Kinematical twist--3 effects in DVCS as a quark spin rotation}
\author{A.V.~Radyushkin}
\address{Physics Department, Old Dominion University, Norfolk, VA 23529, 
USA \\ and \\
Theory Group, Jefferson Lab, Newport News, VA 23606, USA}
\author{C.~Weiss}
\address{Institut f\"ur Theoretische Physik, Universit\"at Regensburg
D--93053 Regensburg, Germany}
\maketitle
\begin{abstract}
We point out that the kinematical twist--3 contributions to the 
DVCS amplitude, required to restore electromagnetic gauge 
invariance of the twist--2 amplitude up to $O(t/q^2 )$, can be understood 
as a spin rotation applied to the twist--2 quark density matrix in the target.
This allows for a compact representation of the twist--3 effects, 
as well as for a simple physical interpretation.
\end{abstract}
\pacs{PACS number(s): 12.38.Bx, 13.60.Fz, 13.60.Le}
}
\thispagestyle{empty}
Various authors have recently studied 
deeply virtual Compton scattering (DVCS) in QCD
beyond the leading twist level 
\cite{Anikin:2000em,Penttinen:2000dg,Belitsky:2000vx,Radyushkin:2000jy,Kivel:2001rb,Radyushkin:2001ap}. 
The main motivation has been the
fact that the twist--2 contribution to the DVCS amplitude alone, 
originally considered in 
Refs.\cite{Ji:1997nm,Collins:1997fb,Radyushkin:1997ki}, 
is not transverse (electromagnetically gauge invariant) even when neglecting
terms of order $t/q^2$. It was shown in 
Refs.\cite{Belitsky:2000vx,Radyushkin:2000jy,Radyushkin:2001ap} that 
transversality to this
accuracy is restored when one includes in the light--cone expansion
certain ``kinematical'' twist--3 contributions which appear from total
derivatives of twist--2 operators. 
Alternatively, these transversality--restoring terms can be represented 
as certain Wandzura--Wilczek type contributions to the general twist--3
skewed parton distributions parametrizing the non-forward matrix elements 
of the relevant light--ray operators beyond the leading twist level
\cite{Anikin:2000em,Penttinen:2000dg,Belitsky:2000vx,Radyushkin:2000jy,Kivel:2001rb,Radyushkin:2001ap}. 
\par
In this Brief Report we point out that the kinematical twist--3
contributions to the DVCS amplitude required by transversality can be 
understood as the result of a certain quark spin rotation applied to the 
twist--2 part of the quark density matrix in the target. This spin rotation 
is necessary in order that the quark density matrix satisfy the free--field 
Dirac equations (up to terms of order $t$) with respect to the ``incoming'' 
and ``outgoing'' quark, 
which is both necessary and sufficient for the tree--level DVCS amplitude to 
be transverse. This observation allows for a very compact representation of 
the twist--3 effects within the non-local light cone expansion,
as well as for a simple physical interpretation.
Notations and conventions in this report closely follow 
Refs.~\cite{Radyushkin:2000jy,Radyushkin:2001ap}.
%
%
\begin{figure}[t]
\begin{center}
\setlength{\epsfxsize}{7.00cm}
\setlength{\epsfysize}{3.62cm}
\epsffile{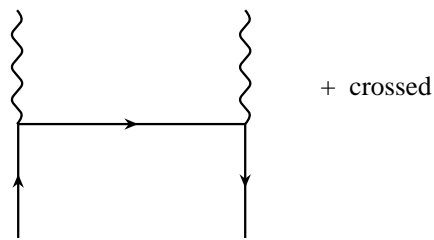}
\end{center}
\caption[]{The virtual Compton amplitude for a free quark.}
\label{fig_free}
\end{figure}
\par
To begin, let us recall how transversality of the tree--level 
Compton amplitude comes about in a theory of free quarks
(of charge unity, for simplicity).
With the vector current given by $J_\mu (x) = \bar\psi (x) \gamma_\mu 
\psi (x)$,
the matrix element of the first contraction of two currents, 
centered at point $X$ and separated by a distance $z$, can be written 
in the form (see Fig.\ref{fig_free})
\be
\lefteqn{
\langle p - r/2| \; i {\rm T} J_\mu (X-z/2) J_\nu (X+z/2) 
\; | p + r/2 \rangle^{\mbox{\scriptsize free q.}} } &&
\nonumber \\
&=& \frac{1}{4}\mbox{tr}\left[ \gamma_\mu \, S(z) \, \gamma_\nu \,
M (z|X) \right] \; + \;
(\mu\leftrightarrow \nu, z \leftrightarrow -z). 
\label{JJ_free}
\ee
Here $S(z) \equiv \hat z/2\pi^2 (z^2 - i0)^2$ is the 
free Dirac propagator in coordinate
space ($\hat z \equiv \gamma_\alpha z_\alpha$), 
and $M$ the density matrix constructed from the 
incoming and outgoing quark wave functions (plane waves),
\beq
M_{ij} (z|X) \;\; = \;\; u_i \bar u_j \;
e^{-i(pz) - i(rX)},  
\label{M_free}
\eeq
where the spinors satisfy $( \hat p + \hat r / 2) \, u = 0$ and
\mbox{$\bar u \, ( \hat p - \hat r / 2 ) = 0$.}
Because of current conservation the expression on the R.H.S.\ 
of Eq.(\ref{JJ_free}) should vanish when contracted with
the derivatives
\beq
\frac{\partial}{\partial z_\mu} - \frac{1}{2} 
\frac{\partial}{\partial X_\mu}, 
\hspace{3em}
\frac{\partial}{\partial z_\nu} + \frac{1}{2} 
\frac{\partial}{\partial X_\nu} .
\eeq
Since the free Dirac propagator satisfies
\beq
\gamma_\mu \frac{\partial}{\partial z_\mu}
S(z) \;\; = \;\; 
\frac{\partial}{\partial z_\nu} S(z)
\gamma_\nu \;\; = \;\;  -i \delta^{(4)} (z),
\eeq
a necessary and sufficient condition for transversality
is that the quark density matrix satisfy the ``left'' and ``right''
Dirac equations
\beq
\left.
\begin{array}{l}
\displaystyle
\left(\frac{\partial}{\partial z_\nu} + \frac{1}{2} 
\frac{\partial}{\partial X_\nu}
\right) \gamma_\nu \, M (z|X)
\\[1ex]
\displaystyle
\left(\frac{\partial}{\partial z_\mu} - \frac{1}{2} 
\frac{\partial}{\partial X_\mu}
\right) M (z|X) \, \gamma_\mu 
\end{array}
\right\}
\;\;  = \;\; 0 .
\label{Dirac_left_right}
\eeq
The free--field density matrix (\ref{M_free}) clearly satisfies
these equation, and thus the R.H.S.\ of Eq.(\ref{JJ_free}) is transverse.
%
%
\begin{figure}[t]
\begin{center}
\setlength{\epsfxsize}{7.00cm}
\setlength{\epsfysize}{5.44cm}
\epsffile{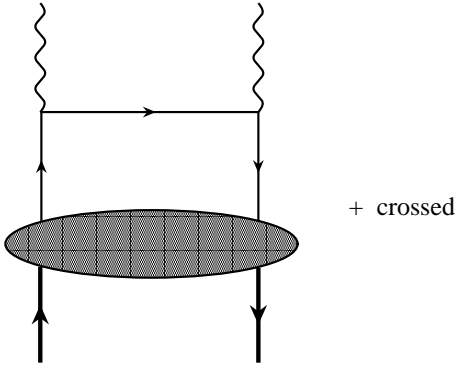}
\end{center}
\caption[]{The twist--2 contribution to the tree--level 
hadronic DVCS amplitude in QCD. The shaded blob represents the
twist--2 part of the quark density matrix in the hadron, 
(\ref{M_twist2}).}
\label{fig_sym}
\end{figure}
\par
Turning now to QCD, the twist--2 contribution to the tree--level 
light--cone expansion of the time--ordered product of two vector
currents has a form analogous to Eq.(\ref{JJ_free}), the
only difference being that the free quark density matrix is
replaced by the twist--2 part of the non-forward matrix element of the
quark density $\bar\psi (X-z/2) \ldots \psi (X+z/2)$ between
hadronic states (see Fig.~\ref{fig_sym}). The latter is defined as
(we suppress the polarization quantum numbers of the hadronic states
for brevity)
\be
\lefteqn{
M_{ij} (z|X)^{\mbox{\scriptsize twist--2}} }
\nonumber \\
&\equiv& \int_0^1 dv 
\left\{ \left( \gamma_\sigma \right)_{ij} \frac{\partial}{\partial z_\sigma}
\langle p - r/2| 
\right.
\nonumber \\
&& 
\left[ \bar\psi (X-vz/2) \, \hat z \, \psi (X+vz/2)
\right]^{\mbox{\scriptsize traceless}}
| p + r/2 \rangle
\nonumber \\
&+& 
\left( \gamma_5 \gamma_\sigma \right)_{ij}
\frac{\partial}{\partial z_\sigma}
\; \langle p - r/2| 
\nonumber \\
&& \left.
\left[ \bar\psi (X-vz/2) \, \hat z \, \gamma_5 \psi (X+vz/2) 
\right]^{\mbox{\scriptsize traceless}}
| p + r/2 \rangle
\right\}
\nonumber \\
\label{M_twist2}
\ee
where the ``traceless'' string operator 
satisfies \cite{Balitsky:1989bk}
\beq
\Box_z \left[ \bar\psi (X-z/2) \, \hat z \,
\psi (X+z/2)
\right]^{\mbox{\scriptsize traceless}} \;\; = \;\; 0 ,
\label{harmonic}
\eeq
and similarly for the operator with Dirac matrix $\hat z \gamma_5$.
Only the chirally even part of the density matrix has been included 
in Eq.(\ref{M_twist2});
the chirally odd (transversity) part with Dirac structure 
$\sigma_{\alpha\beta}$ would enter the DVCS amplitude only at
power--suppressed level.
Noting that the coefficient function for the twist--2 
contribution in QCD is the same as that in the free theory --- the free 
quark propagator, $S(z)$ --- it is evident that transversality would 
again require that the quark density matrix satisfy the free Dirac 
equation. Substituting Eq.(\ref{M_twist2}) in Eqs.(\ref{Dirac_left_right}) 
one finds
\be
\lefteqn{
\left.
\begin{array}{r}
\displaystyle
\left(\frac{\partial}{\partial z_\nu} + \frac{1}{2}
\frac{\partial}{\partial X_\nu}
\right) \gamma_\nu \, M (z|X)^{\mbox{\scriptsize twist--2}}
\\[2ex]
\displaystyle
\left(\frac{\partial}{\partial z_\mu} - \frac{1}{2}
\frac{\partial}{\partial X_\mu}
\right)  M (z|X)^{\mbox{\scriptsize twist--2}} \, \gamma_\mu
\end{array}
\right\}
} &&
\nonumber \\
&=& \left[ \Box_z + \frac{1}{2} 
\left( \frac{\partial}{\partial X} \frac{\partial}{\partial z} \right)
\pm \frac{1}{2} 
\left( \frac{\partial}{\partial X} \sigma 
\frac{\partial}{\partial z} \right)
\right] \int_0^1 dv
\nonumber \\
&& \times 
\langle p - r/2| \left[ \bar\psi (X-vz/2) \hat z \psi (X+vz/2)
\right]^{\mbox{\scriptsize traceless}}
\nonumber \\
&& \phantom{\times} | p + r/2 \rangle
\nonumber \\
&+& \ldots ,
\label{Dirac_twist2}
\ee
where the ellipsis stands for the corresponding term with the operator 
$\bar\psi \hat z \gamma_5\psi$, which we have not written out for brevity.
The first term in the bracket vanishes because the string operator satisfies
Eq.(\ref{harmonic}). The second term, involving the contracted derivatives,
is proportional to $t \equiv r^2$ or $m^2$ (the target mass squared) and can 
be dropped if we are interested in the DVCS amplitude up to terms of 
order $t/q^2$ or $m^2/q^2$. The third term, however, involving the structure
\beq
\left( \frac{\partial}{\partial X} \sigma 
\frac{\partial}{\partial z} \right) \;\; \equiv \;\;
\frac{\partial}{\partial X_\alpha} \sigma_{\alpha\beta} 
\frac{\partial}{\partial z_\beta} ,
\label{problematic}
\eeq
where $\sigma_{\alpha\beta} \equiv (1/2) [\gamma_\alpha , \gamma_\beta ]$,
is not proportional to either $t$ or $m^2$, and thus cannot be neglected 
in DVCS kinematics. Thus, the twist--2 density matrix (\ref{M_twist2}) does 
not satisfy the free Dirac equation even when neglecting terms
proportional to $t$ or $m^2$. This is the reason why the twist--2 
contribution to the DVCS amplitude is not transverse up to terms
of order $t/q^2$ or $m^2/q^2$, as has been observed within 
various different approaches in
Refs.\cite{Anikin:2000em,Penttinen:2000dg,Belitsky:2000vx,Radyushkin:2000jy,Kivel:2001rb,Radyushkin:2001ap}.
\par
\begin{figure}[t]
\begin{center}
\setlength{\epsfxsize}{7.00cm}
\setlength{\epsfysize}{7.41cm}
\epsffile{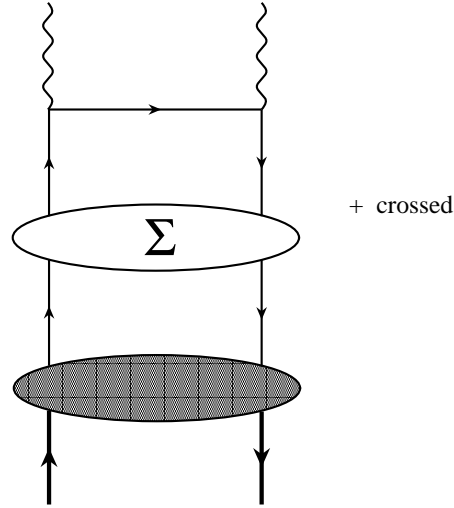}
\end{center}
\caption[]{Diagrammatic representation of the quark spin rotation, 
Eqs.(\ref{M_rot}) and (\ref{JJ_rot}), as an intermediate step between the 
twist--2 quark density matrix in the hadron and the free--quark virtual
Compton amplitude.}
\label{fig_rot}
\end{figure}

One may now ask how the twist--2 part of the quark density matrix 
(\ref{M_twist2}) could be modified such that it satisfies
the free--field Dirac equations up to terms of order $t$ or $m^2$.
Since the problematic term in Eq.(\ref{Dirac_twist2}) involves the 
Dirac matrix $\sigma_{\alpha\beta}$, it seems natural to look for a
transformation in the form of a quark spin rotation. 
Since we require that the transformation matrix should reduce to unity
for forward matrix elements, the only possibility (up to a variable factor 
in the exponent) is a transformation of the form
\be
\Sigma\,(z/2) &\equiv& \exp \left[ -\frac{1}{4} \left(
z \sigma \frac{\partial}{\partial X} \right) \right] ,
\label{S}
\\
\left( z \sigma \frac{\partial}{\partial X} \right) 
&\equiv&
z_\alpha \sigma_{\alpha\beta} \frac{\partial}{\partial X_\beta} ,
\ee
which satisfies
\beq
\Sigma\,(z/2)^{-1} \;\; = \;\; \Sigma\,(-z/2) .
\eeq
It is easy to see that a density matrix with the desired properties is 
obtained by the following transformation:
\be
\lefteqn{
M_{ij} (z|X)^{\mbox{\scriptsize rot}} \;\; \equiv \;\; \int_0^1 dv \; 
\left\{ \phantom{\frac{\partial}{\partial z}}
\right. } &&
\nonumber \\
&& 
\left[ \Sigma\,(\bar v z/2 ) \, \gamma_\sigma \, 
\Sigma\,(\bar v z/2 ) \right]_{ij}
\frac{\partial}{\partial z_\sigma}
\; \langle p - r/2| 
\nonumber \\
&& 
\left[ \bar\psi (X-vz/2) \, \hat z \, \psi (X+vz/2)
\right]^{\mbox{\scriptsize traceless}}
| p + r/2 \rangle
\nonumber \\
&+& 
\left[ \Sigma\,(\bar v z/2 ) \, \gamma_5 \gamma_\sigma \, 
\Sigma\,(\bar v z/2 ) \right]_{ij}
\frac{\partial}{\partial z_\sigma}
\; \langle p - r/2| 
\nonumber \\
&& \left.
\left[ \bar\psi (X-vz/2) \, \hat z \gamma_5 \, \psi (X+vz/2) 
\right]^{\mbox{\scriptsize traceless}}
| p + r/2 \rangle
\right\} ,
\nonumber
\\
\label{M_rot}
\ee
where $\bar v \equiv 1 - v$. Indeed, by a straightforward calculation, 
making use of the explicit form of the matrix exponential
\be
\Sigma\,(\bar v z/2) &=&
\cosh \left[ \frac{\bar v}{4} 
\left( z \frac{\partial}{\partial X} \right) \right]
\nonumber \\
&-& \frac{\displaystyle \sinh \left[ \frac{\bar v}{4} 
\left( z \frac{\partial}{\partial X} \right) \right]}
{\displaystyle \left( z \frac{\partial}{\partial X} \right)}
\left( z \sigma \frac{\partial}{\partial X} \right)
\nonumber \\
&+& \mbox{terms} \;\; \left( 
\frac{\partial}{\partial X} \frac{\partial}{\partial X}
\right) ,
\ee
one can convince oneself that the ``rotated'' twist--2 density matrix
(\ref{M_rot}) satisfies the free--field left-- and right Dirac equations
(\ref{Dirac_left_right}) up to terms involving
contracted derivatives of the string operator of the form
\beq
\left( \frac{\partial}{\partial z} \frac{\partial}{\partial X} \right) ,
\hspace{3em}
\left( \frac{\partial}{\partial X} \frac{\partial}{\partial X} \right)
\eeq
which have matrix elements of order $t$. Diagrammatically, the spin rotation
of the twist--2 density matrix may be represented as in Fig.\ref{fig_rot},
as a kind of ``evolution kernel'' acting between the twist--2 matrix
element and the free quark propagator in the coefficient function.
Note that since in the matrix element
the total derivative operator just turns into the momentum 
transfer, $\partial / \partial X_\alpha = i r_\alpha$, the ``rotated'' 
density matrix does not involve any information beyond the basic twist--2 
matrix elements, as parametrized by the twist--2 skewed distributions.
%
%
\par
The contribution to the light--cone expansion generated by the ``rotated'' 
density matrix (\ref{M_rot}) may be written in the form
\be
&& \frac{1}{4}\mbox{tr}\left[ \gamma_\mu \, S(z) \, \gamma_\nu
\, M (z|X)^{\mbox{\scriptsize rot}} \right] \; + \;
(\mu\leftrightarrow \nu, z \leftrightarrow -z). 
\nonumber \\
&=& \frac{1}{2 \pi^2 z^4} \int_0^1 dv \; \left\{ 
\frac{1}{4} \mbox{tr}
\left[ \Sigma\,(\bar v z/2 ) \, \gamma_\mu \, \hat z \, \gamma_\nu 
\, \Sigma\,(\bar v z/2 )
\gamma_\sigma \right] 
\right. 
\nonumber \\
&& \times \frac{\partial}{\partial z_\sigma}
\; \langle p - r/2| 
\nonumber \\
&& \left[ \bar\psi (X-vz/2) \, \hat z \, \psi (X+vz/2) 
\right]^{\mbox{\scriptsize traceless}} | p + r/2 \rangle
\nonumber 
\\
&+& \frac{1}{4} \mbox{tr}
\left[ \Sigma\,(\bar v z/2 ) \, \gamma_\mu \, \hat z \, \gamma_\nu 
\, \Sigma\,(\bar v z/2 ) \,
\gamma_5 \gamma_\sigma
\right]
\nonumber \\
&& 
\times \frac{\partial}{\partial z_\sigma}
\; \langle p - r/2| 
\nonumber \\
&& \left.
\left[ \bar\psi (X-vz/2) \, \hat z \gamma_5 \, \psi (X+vz/2) 
\right]^{\mbox{\scriptsize traceless}} | p + r/2 \rangle
\right\}  
\label{Pi_transverse_S}
\nonumber \\
&+& (\mu\leftrightarrow \nu, z \leftrightarrow -z),  
\label{JJ_rot}
\ee
where we have substituted the explicit form of the free coordinate--space 
quark Green function. In this expression the spin rotation appears
as acting on the coefficient functions rather than the matrix elements;
this is of course a matter of interpretation. One can easily verify that 
Eq.(\ref{JJ_rot}) is identical to the expression in Eq.(3.51) of 
Ref.\cite{Radyushkin:2001ap}, which was obtained by making repeated use
of an string operator identity following from the QCD equations of motion
and neglecting quark--gluon operators.
Thus, we see that the ``kinematical'' twist--3 contributions to the
non-forward light--cone expansion (and consequently to the DVCS amplitude) 
can be understood as the effect of a quark spin rotation applied to the 
twist--2 part of the quark density matrix.
\par
We have chosen here to present the spin rotation at the level of non-forward
matrix elements of the quark density, as relevant to DVCS. It is clear, 
however, that this representation of ``kinematical'' twist--3 contributions 
in the light--cone expansion is valid at the operator level, and thus can be 
applied also in the context of other exclusive reactions where the relevant 
matrix elements are of distribution amplitude type.
\par
This work was supported by the U.S. 
Department of Energy under contract
DE-AC05-84ER40150 under which the Southeastern
Universities Research Association (SURA)
operates the Thomas Jefferson National Accelerator
Facility (Jefferson Lab),  by the Alexander  
von Humboldt Foundation, 
by the Deutsche 
Forschungsgemeinschaft (DFG) and  by
the German Ministry of Education and Research (BMBF).
\end{document}